\documentstyle[preprint,prc,aps]{revtex}
\begin{document}
\preprint{
\rightline{\vbox{\hbox{\rightline{MSUCL-1026}} 
\hbox{\rightline{nucl-th/9604036}}}}
         }
\title{Collision Broadening of Rho Meson in a Dropping Mass Scenario}
\author{K. L. Haglin\cite{myemail}}
\address{
National Superconducting Cyclotron Laboratory, Michigan
State University\\
East Lansing, Michigan 48824--1321, USA}
\date{\today}
\maketitle 
\begin{abstract} 
Vector mesons containing light quarks are thought to have their masses 
reduced in dense nuclear matter, sacrificing some of their energy to the 
scalar field which becomes appreciable at finite baryon density.  Model
calculations find masses which fall by a couple tens of percents in normal 
nuclear matter, and by several hundred MeV in dense matter.  We estimate the 
collision rate for rho mesons in such a scenario and at finite temperature.  
Compared to its free-mass value, the collision rate changes by nearly a factor
of two both above and below, depending on the density.  This collision 
broadening effect could be important for estimates of low-mass dilepton 
production in heavy-ion collisions.
\end{abstract}

Energetic heavy-ion collisions produce matter far the ground state and provide
an exploratory workplace for ideas about strong interaction physics modifying
particle properties.  Restoration of chiral symmetry is an example which 
could have dramatic effects on the overall dynamics of the collision by 
changing particle masses (Brown-Rho scaling)\cite{br91}, cross sections, 
propagation and decay rates.  Several issues remain unresolved. For instance, 
resonance spectral densities could become degenerate with their chiral 
partners, they could remain separate but mix very strongly or they could 
disappear altogether by effectively melting into continua.  One popular 
picture is to couple nucleons to scalar and vector fields and to couple light
vector mesons to the scalar field in a self-consistent manner.  This 
Walecka-type of approach leads to masses that drop nearly universally with 
increasing density\cite{ab93} and presumably become degenerate, or nearly so.
However, any deviation from free space behavior is brief and unlikely 
observable in terms of hadronic signals since it is masked behind layers of 
complicated dynamics and spacetime evolutionary information redistribution.  
On the other hand, dilepton signals can in principle provide a clear snapshot
of the isoscalar and neutral-isovector vector mesons since the electromagnetic 
quanta couple weakly to the hadronic medium.  Final spectra exhibit resonance 
structure which, in the low-mass region, includes rho and omega mesons.   The 
natural width of the omega being 8.43 MeV, dictates a lifetime which is too 
long for decay in the high density part of the evolution and so one turns to 
the rho meson for opportunity. 

Recent dielectron measurements have reported an enhancement in the low-mass 
region for 200 GeV/u S+Au collisions at the Super Proton Synchrotron as 
compared with expected yields from hadronic decays\cite{ga95}.  The 
enhancement was most clear around 400 MeV mass and one possible explanation 
among those proposed\cite{lk95,ds96,kh96,jw96} is that the rho mass drops 
in the early, high density stage of the evolution and provides a component to 
the dielectron signal much below its free 
mass\cite{lk95}.  Later stages of evolution support lower densities and 
consequently higher rho meson masses and will provide dielectron components 
all the way up to a free-space rho distribution.  The question we address 
here is the relative importance of collision broadening in such a picture.

To further motivate and quantify the dropping mass scenario, we imagine a 
system of baryons, pseudoscalar, vector and axial-vector mesons at 
temperature $T$ and baryon density $\rho_{B}$, and we look to models
of the in-medium hadronic properties.  They include QCD sum-rule 
approaches\cite{dei90,hl92}, effective field theories based largely on 
particle-hole polarization effects\cite{sh94}, quark models\cite{st94} 
and mean-field or Walecka-type models\cite{sw86}.   A trend they seem 
to share is the dropping mass at finite density.  In the Walecka-type 
approach, the nucleons couple to vector and scalar fields while the light 
vector mesons couple only to the scalar field since they are composed of quark
and antiquark giving counterbalancing contributions.  For given baryon 
density, the energy density $\varepsilon$ is a function of the scalar field 
$\bar\sigma$.  By locating the extrema of $\partial \varepsilon/\partial 
\bar\sigma = 0$, one establishes the strength of the scalar field.  Then the 
constituent quark model\cite{st94} together with Weinberg sum rules\cite{sw67}
for relating $\rho$ and $a_{1}$ masses give, for example, 
\begin{eqnarray}
m^{*}_{N} = m_{N} - g_{S}\bar\sigma , \quad
m^{*}_{\rho} \approx m_{\rho} - (2/3)g_{S}\bar\sigma, \quad
m^{*}_{a_{1}} \approx m_{a_{1}} - (2\sqrt{2}/3)g_{S}\bar\sigma,
\end{eqnarray}
where the starred quantities are in-medium values and $g_{S}$ is a coupling 
constant determined by balancing the repulsive vector and attractive scalar 
influences in order to reproduce bulk properties of nuclear matter. Ignoring 
medium modifications to pions and kaons then leads to a picture of baryon, 
vector and axial-vector meson masses all dropping with increasing density.  
At this stage, choosing a specific effective rho mass determines the scalar 
field which in turn, determines the density.  For given scalar field, all 
other hadron masses are then determined.  It is therefore a well defined 
question in this model to ask for the collision rate at fixed $T$ as a 
function of effective rho mass from $m_{\rho}$ = 770 MeV down to the two-pion
threshold. 

Collision rates are most conveniently estimated in a fireball model and from
kinetic theory rather than some sort of dynamic model.  This approach has 
been carried out to a rather extensive level for $\rho$, $\omega$ and 
$\phi$-mesons interacting with pseudoscalar and light vector 
mesons\cite{kh95}.  Interactions were parametrized according to the 
simplest lagrangians respecting proper symmetries and then calibrated to 
give the free-space decays.  Tree level Feynman graphs were enumerated and 
computed all while checking unitarity bounds and including form factors to 
account for the composite nature of the particles.  Collision rates for 
$\rho$ and $\omega$ were found to be numerically similar and ranged from 
40--120 MeV as the temperature varied from 150--200 MeV.  The completeness of 
the calculation was crucial to getting the $\phi$ result of 10--25 MeV since 
there are no strong resonances.  But for $\rho$ and $\omega$, even with many 
nonresonant contributions, the resonances dominated the cross sections and 
consequently, the rates.  This simplifies things enormously since we can be 
satisfied with Breit-Wigner parametrizations for the cross sections and 
eliminate the need for all that complexity.

Consider the system of an hadronic mixture of pions, kaons, rho mesons 
and nucleons.   Introduction of eta, omega or higher mass mesons does not 
change the results significantly.  Give them equilibrium distributions 
$f(p,\mu)$, having the flexibility of a chemical potential $\mu$.  It is 
probably true that an equilibrated system is not fully reached for all of 
these constituents in heavy-ion collisions, but for first estimates it is the 
most reasonable thing to assume.   Rho mesons scatter vigorously with pions 
through the $a_{1}(1260)$ resonance and with kaons through the $K_{1}(1270)$, 
both of which are axial-vectors and carry isospin 1 and 1/2 respectively.  
The $K_{1}$ mass is assumed to scale just like the $a_{1}$'s.  Rho-rho
and rho-nucleon scattering give contributions which are very much
smaller, so we will restrict attention only to pions and kaons.

Ignoring final state suppression or enhancements and approximating with
classical distributions, the average scattering rate is 
\begin{eqnarray}
\overline{\Gamma}_{\rho}^{\rm \, coll} &=& \sum\limits_{i} 
\exp(\mu_{i}/T){g_{\rho}g_{i}\over
n^{*}_{\rho}}{T^{2}\over (2\pi)^{4}} \int\limits_{z_{\rm min}}^{\infty} dz \, 
\lambda(z^2T^2,m_{\rho}^{*\,2},m_{i}^{2})
\sigma_{\rho\, i}(s)
\label{eq:rate}
\end{eqnarray} 
where the sum runs over $i=\{\pi, K\}$, the $g$'s are degeneracies,
$n^{*}_{\rho}$ is the particle density of a thermal population of rho
mesons having mass $m_{\rho}^{*}$ and zero chemical potential,
$z_{\rm min} = (m^{*}_{\rho}+m_{i})/T$ and $\lambda$ is the usual kinematic 
function.  For the cross sections
we take Breit-Wigner functions and we compute energy dependent widths.
As the masses drop, phase space changes which naturally has modifying effects
on the widths and cross sections.  The simplest lagrangian respecting both 
gauge invariance and minimal pion derivatives is chosen for the 
axial-vector--vector--pseudoscalar (A-V-P) interactions.  It has been used 
for photon\cite{es92,kh94} and dilepton\cite{kh96} production estimates and 
is known to be reliable.  The form of the interaction and the resulting 
strong decay rates are
\begin{eqnarray}
L_{AVP} &=& -g_{AVP}A_{\mu}\partial_{\nu}P\,\,\,V^{\mu\nu}\\
\Gamma_{A\to\,VP} &=& {g_{AVP}^{2} \over 24\pi\,m_{A}^{2}}
\,|\vec{p}\,|\left[2(p_{V}\cdot p_{P})^{2}+m_{V}^{2}(m_{P}^{2}+
\vec{p}^{\,\,2})\right]
\label{eq:decay}
\end{eqnarray}
where $V^{\mu\nu}=\partial^{\mu}V^{\nu}-\partial^{\nu}V^{\mu}$ and 
$\vec{p}\,$ is the center of mass momentum of the decay products.
Cross sections are taken to be
\begin{eqnarray}
\sigma_{V\,i} &=& {4\pi \over \vec{p}^{\,\,2}} 
\left[{  m_{R}^{2}\Gamma_{R\to V\,i}^{2} \,
B_{R\to V\,i}^{2}
\over (s-m_{R}^{2})^{2} + (m_{R}\Gamma_{R}^{\rm \, tot})^{2} }\right]
\end{eqnarray}
since the spin factor ratio is 1 for both $\pi\,\rho$ and $K\,\rho$ 
scattering through $a_{1}$ and $K_{1}$ resonances, respectively. The 
scattering rate of Eq.~(\ref{eq:rate}) is now a function of 
temperature, chemical potentials and hadron masses.  Here we assume chemical 
equilibrium with a single potential $\mu_{\pi}$, and with the rho chemical 
potential determined solely from it.  As mentioned previously, choosing a 
value of effective rho mass determines the effective masses for the other
hadrons as well.

In Fig.\ref{fig:one} we show the rate at $T$ = 150 MeV as a function of 
the rho mass choosing two values of pion chemical potential (0 and 130 MeV).
In the free-mass extreme and zero chemical potential case,
we can check against the rho collision rate of $\sim$ 40 MeV computed in 
Ref.~\cite{kh95} with more sophisticated means.  The two are consistent 
to within 5$\,$\%, so the present approach is quite satisfactory.  As the rho
mass drops from its free-space value of 770 MeV, the collision rate rises to a 
maximum of 75 MeV for a mass around 500 MeV.  From here it drops to 40 MeV at
400 MeV mass and to 20 MeV at 300 MeV mass.  Thus, for the lightest rho 
mesons considered here, the collision broadening is clearly not important.  On 
the other hand, for intermediate rho masses the collision rates reach
75--175 MeV depending on the pion chemical potential.  Under these 
circumstances, collision broadening is most certainly important.

We recompute at a higher temperature of 200 MeV and show the results
in Fig.~\ref{fig:two}.   The rate for a free mass and zero
chemical potential is $\sim$ 140 MeV.  This is about 15\% higher than the
result of Ref.~\cite{kh95} and due to the differences in the cross sections 
used here.   The peak in the rate seen at low temperature is largely smeared 
out since the average scattering energies are higher here and are therefore 
able to access the resonance peaks more easily.  The key elements for 
determining the rate are the positions and widths of the resonances compared 
to the average $\sqrt{s}$ for individual scatterings.   In this higher 
temperature result the resonances are accessed equally well down to 500 MeV 
mass and from there start to drop off the tail of the resonance distributions.
At $T$= 150 MeV, on the other hand, as the mass drops the resonance is
accessed more favorably and it is in fact narrower and more noticeably 
spiked, thereby giving a boost to the rate.
   
We have computed the collision rate of the rho meson in hot and dense 
hadronic matter as a function of the effective rho mass or alternatively and
equivalently, as a function of density.  At $T$ = 150 MeV and
low densities the rate is $\Gamma^{\rm\, free\,\,mass}_{\rho} \sim$ 40 MeV.  
As the density increases and the mass drops, the rate rises to about twice 
$\Gamma^{\rm\, free\,\,mass}_{\rho}$, peaks and then drops to about half 
$\Gamma^{\rm\, free\,\,mass}_{\rho}$ near the two-pion threshold.  Since mass 
resolution in the current dilepton experiments is roughly 10\%, rho meson 
collision rates of a few tens of MeV are not too important since the two 
effects (resolution and collision broadening) add in quadrature.  At higher 
temperatures we have seen the rates rise to more than 100 to 250 MeV depending
if a chemical potential is introduced.  Both values are significant and 
suggest that collision broadening could be quite important for model 
calculations of low-mass dileptons when quantitative interpretation of 
experiment is attempted.

\section*{Acknowledgments}

I thank C.M. Ko for asking the questions which lead to this study.
This work was supported by the National Science Foundation under grant 
number PHY-9403666.

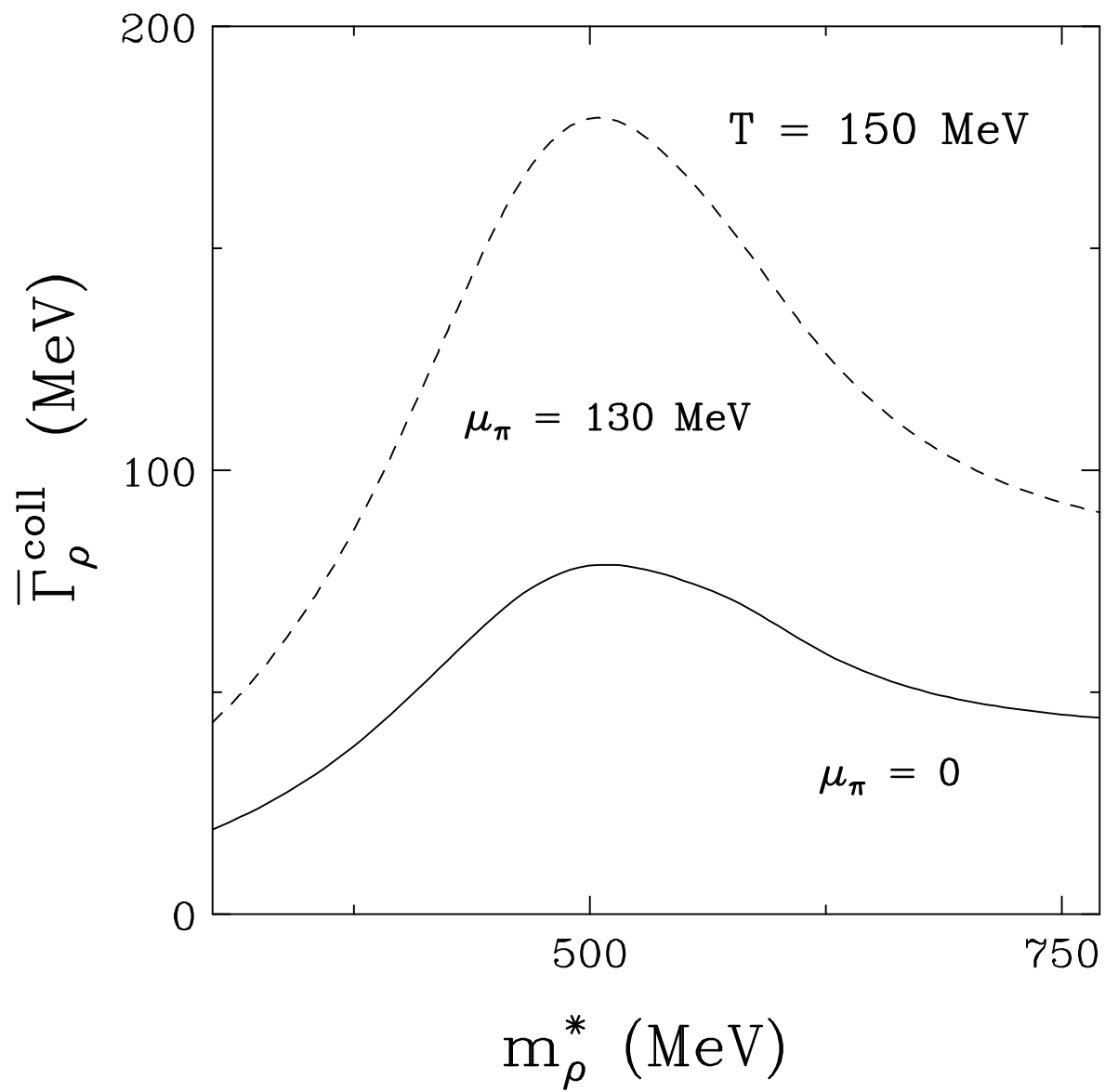
\begin{figure}
\caption{Collision rate for $\rho$ mesons at fixed temperature
$T$ = 150 MeV as a function of the effective rho mass.
Two different values of pion chemical potential are used.  The
solid and dashed curves come from $\mu_{\pi}$ = 0 and 130 MeV, respectively.}
\label{fig:one}
\end{figure}
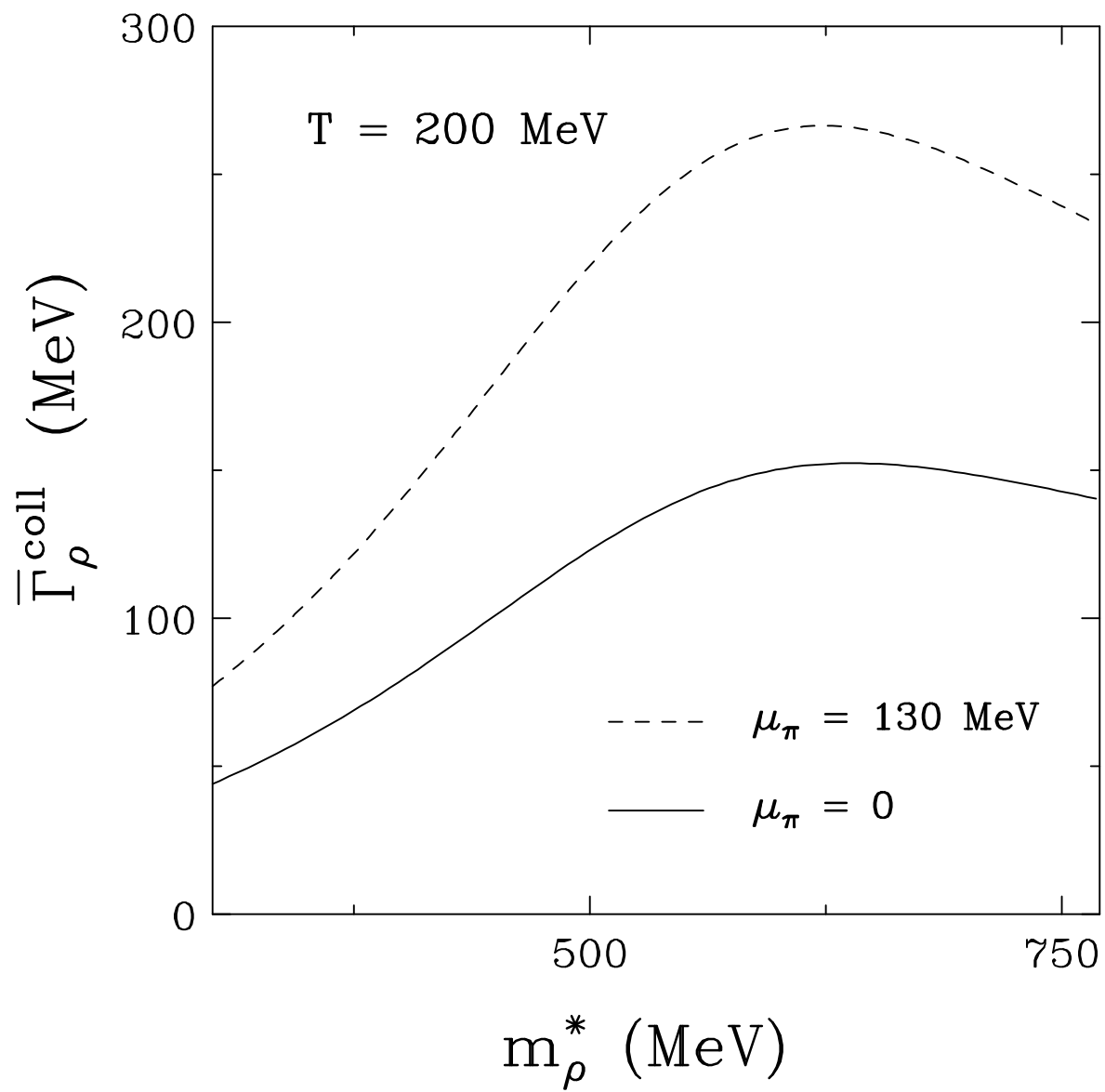
\begin{figure}
\caption{Same as previous figure but with $T$ = 200 MeV.}
\label{fig:two}
\end{figure}
\end{document}